# Coupled magnetic and elastic properties in LaPr(CaSr)MnO manganites

G.G. Eslava[1], F. Parisi[2,3], P.L. Bernardo[1], M. Quintero[2,3], G. Leyva[2,3], L.F. Cohen[4] and L. Ghivelder[1,*]

[1] Instituto de Física, Universidade Federal do Rio de Janeiro, Rio de Janeiro, RJ, 21941-972, Brazil

[2] Escuela de Ciencia y Tecnología, Universidad Nacional de General San Martín, Buenos Aires, Argentina.

[3] Departamento de Física de la Materia Condensada, Comisión Nacional de Energía Atómica, Buenos Aires, Argentina

[4] Blackett Laboratory, Imperial College, London, SW7 2AZ, UK

abstract

We investigate a series of manganese oxides, the $La_{0.225}Pr_{0.4}(Ca_{1-x}Sr_x)_{0.375}MnO_3$ system. The $x$ = 0 sample is a prototype compound for the study of phase separation in manganites, where ferromagnetic and charge ordered antiferromagnetic phases coexist. Replacing $Ca^{2+}$ by $Sr^{2+}$ gradually turns the system into a homogeneous ferromagnet. Our results show that the material structure plays a major role in the observed magnetic properties. On cooling, at temperatures below ~100 K, a strong contraction of the lattice is followed by an increase in the magnetization. This is observed both through thermal expansion and magnetostriction measurements, providing distinct evidence of magneto-elastic coupling in these phase separated compounds.

* corresponding author (Luis Ghivelder); e-mail: luisghiv@if.ufrj.br

## 1 - Introduction

The study of manganese based perovskites, known as manganites, has fascinated the scientific community for the last two decades. This interest was triggered by the discovery of the colossal magnetoresistance effect [1], a huge reduction of the electrical resistivity on application of a magnetic field, which was explained in terms of the interplay between the double exchange mechanism and quenched disorder [2]. In addition, the competition between very different phases with similar ground state energies, namely an insulating charge ordered antiferromagnetic (CO-AFM) and a metallic ferromagnetic (FM) state, leads to complex phase diagrams [3], and to the occurrence of a phase separated state [4,5], where both phases can spontaneously coexist in meso- or nanoscopic scales [6,7,8]. Phase separation (PS) in manganites has became a very important topic of investigation in the physics of strongly correlated electron systems, and a similar phenomenology is found in the high Tc cuprates [9], particularly in the underdoped regime, as well as in magnetocaloric materials, such as $Gd_5Ge_4$ [10].

A prototype compound to investigate the phenomenon of phase separation in manganites is $La_{5/8-y}Pr_yCa_{3/8}MnO_3$, which has been thoroughly studied in the literature [11,12,13,14]. It is a mixture of the optimized colossal magnetoresistive compound $La_{0.625}Ca_{0.375}MnO_3$, with the Pr based system, $Pr_{0.625}Ca_{0.375}MnO_3$, known to have a robust CO-AFM state at low temperatures. The intermediate mixture, $La_{0.225}Pr_{0.40}Ca_{0.375}MnO_3$, becomes charge ordered and highly insulating below 230 K, and orders antiferromagnetically below 180 K. Phase separation manifests itself at lower temperatures, below ≈ 80 K, where FM metallic clusters which nucleate within the CO-AFM matrix yield a sizable magnetization even in low applied magnetic fields.

In this work we performed additional doping to the above mentioned phase separated compound, $La_{0.225}Pr_{0.40}Ca_{0.375}MnO_3$, by replacing the alkaline-earth $Ca^{2+}$ with larger $Sr^{2+}$ ions. This substitution introduce structural distortions that gradually increases the exchange coupling [15], thus promoting the FM phase at low temperatures, eventually leading to a homogeneous FM ground state. Our objective is to correlate the macroscopic magnetic response of the system with its structural behavior, obtained from thermal expansion and magnetostriction measurements. The results clearly show that the formation of FM clusters on cooling occurs at temperatures just below a strong lattice contraction of the material, giving evidence of the coupling between structural and magnetic degrees of freedom in phase separated manganites.

**2 - Experimental**

Polycrystalline samples of $La_{0.225}Pr_{0.4}(Ca_{1-x}Sr_x)_{0.375}MnO_3$, with $0 \leq x \leq 1$, were synthesized by a citrate-nitrate decomposition method [16], using 99.9% purity reactants (oxides and soluble salts). After the mixed solution was dried at 100º C, thermal treatments were made at 500 °C for 5 hours and at 1400 °C for 16 hours. The powder was compacted into pellets of 5x2x2 mm3 using a pressure 0.8 t/cm$^2$. The pellets were finally sintered at 1400ºC for 2 hours. The resulting average grain size is ~ 2 μm. Magnetization and specific heat measurements were made with a PPMS system (manufactured by Quantum Design, Inc.), the former using a VSM magnetometer. Thermal expansion and magnetostriction measurements were made with a silver based capacitance dilatometer [17], fitted in the PPMS sample chamber. High resolution x-ray powder diffraction (XPD) measurements were conducted at the Brazilian Synchrotron Light Laboratory (LNLS). The polycrystalline samples used for the x-ray data were ground and sieved to reject grains larger than 5 μm. A wavelength of 1.7614 Å was employed and a Ge(111) crystal analyzer used for detection.

## 3 - Results and Discussion

Figure 1 shows the room temperature x-ray diffraction intensity of $La_{0.225}Pr_{0.4}(Ca_{1-x}Sr_x)_{0.375}MnO_3$, with $x$ = 0.05. Rietveld refinements were carried out with the FULLPROF program, with an orthorhombic *Pnma* space group. The reliability factors of the Rietveld analysis [18] are $R_{Bragg}$ = 9.04, $R_F$ = 9.81, and $\chi^2$ = 1.57. The inset shows an enlarged plot of specific areas of the diffractogram, so that the calculated and observed profiles are clearly visible. The lower trace in the main panel is a plot of the difference. Similar results were obtained for all other samples. All diffraction peaks were accounted for, with no impurity phases within the precision of the measurements, confirming the good quality of the samples under study. Relevant microscopic parameters extracted from the XPD analysis are the Mn-O distance and the Mn-O-Mn bond angle, plotted in Fig. 2 for a complete range of Sr doping, $0 \leq x \leq 1$. Microscopically, adding Sr into the compounds increases the mean ionic radius of the A site, $\langle r_A \rangle$, which in turn increases the tolerance factor $t$ [19], a geometrical factor which quantifies the lattice distortion ($t$ = 1 for the cubic structure). If we analyze the behavior of the tolerance factor as a function of the Sr content in the series, we find that $t$ = 0.919 in the fully Ca doped sample, and increases to 0.937 in the fully Sr doped compound. Figure 2 shows how the calculated values of $t$ and $\langle r_A \rangle$ [20] correlate with the Sr content in the samples. A larger tolerance factor implies a less distorted lattice, which favors the double exchange mechanism, responsible for the FM interaction in manganites. The increase of the Mn-O-Mn bond angle for higher Sr doping is consistent with the enhanced FM character of the compound, as evident in the magnetic data which follows.

Figure 3a shows the temperature dependence of the magnetization of the samples with a Sr doping $x \leq 0.4$, measured with a moderate field H = 1 T. Samples with x>0.4 display qualitatively the same behavior as for the x=0.4 . It is readily observed that adding Sr favors the FM character of the compounds. Compounds with higher Sr content are nearly homogeneous ferromagnets at low temperatures while those with low Sr content present evidence of charge ordering at intermediate temperatures, and a behavior consistent with phase separation in a wide temperature range [6,12,21]. Figure 3b shows the influence of different applied magnetic fields on the magnetization of the sample with $x$ = 0.05. Comparing Figs. 3a and 3b it is interesting to note that applying magnetic fields in the range between 0.2 T and 5 T on a low doped sample has a similar effect on the magnetic properties as increasing $\langle r_A \rangle$ at a fixed H value. The results of Fig. 3 are a clear indication of the interplay between the structure and magnetic properties of the system: chemical doping, which yield a less distorted structure, is equivalent to applying an increased magnetic field on a low doped sample. This is expected on the basis of the double exchange mechanism, and agrees with previous report of the effect of Sr substitution at the Ca site in phase separated manganites [15].

As observed in Fig. 3a, the compound with Sr content x = 0.05 displays clear signatures of phase separation [21], and contains a larger fraction of the low temperature FM phase than the x = 0 undoped sample. Therefore, $La_{0.225}Pr_{0.4}(Ca_{0.95}Sr_{0.05})_{0.375}MnO_3$ is an appropriate compound to study the coupling between magnetic and elastic properties. Evidence of the interplay between structure and magnetism can be obtained through dilatometry measurements. Figure 4 shows results of thermal expansion and low field magnetization. The specific heat curve is also plotted in the same graph. At T ≈ 220 K there is an anomaly associated with the CO transition, clearly identified in all measurements: thermal expansion, magnetization and specific heat. The latter also shows the AFM transition at ≈ 185 K. An expansion of the compound on cooling related to the cooperative Jahn-Teller (JT) distortion is observed in a small temperature window just below the CO transition. With further cooling the JT distortion eventually disappears, giving rise to electron mobility and the formation of long range FM clusters. This is associated with a rapid contraction of the lattice observed below 100K. The inset of Fig. 4 shows the temperature derivatives of the magnetization and of the relative thermal expansion. The latter, known as the thermal expansion coefficient, shows a distinct peak, at ~ 85 K, which coincides with the onset of FM on cooling as observed in the dM/dT data. This is a direct evidence of the coupling between the structural and magnetic properties: the lattice contraction gives rise to the formation of the FM phase. Surprisingly, no anomaly is observed in the specific heat associated with this increase of ferromagnetism and the lattice contraction, indicating that it is not a thermodynamic phase transition. A similar scenario, the absence of an entropy related anomaly in a superconducting perovskite cuprate, was previously used as evidence to support the absence of a phase transition [22].

The magneto-elastic coupling can also be probed through magnetostriction measurements. In Fig. 5 we shown the results with applied magnetic fields up to H = 4 T at T = 100 and 70 K. It is observed that the application of magnetic field promotes simultaneously an increase of the magnetization and a contraction of the sample's dimensions. At 100 K, above the peak in thermal expansion coefficient (inset of Fig. 4), this transformation is reversible. On the other hand, for temperatures below the peak in the thermal expansion coefficient, the field induced transformation is irreversible: when the field is lowered the compound retains its contraction and remains in the FM state.

The overall low temperature magnetic and elastic behavior of the x=0.05 sample can be summarized as follows: on cooling with low fields (Fig. 4) a fraction of the sample transforms to a less distorted structure (at ≈ 80 K), followed by the appearance of a long range FM state at temperatures just below the structural changes. It is assumed that the portion of the sample that becomes FM is the same portion that experiences the volume contraction. This enhanced contraction occurs because the JT distortions associated with the higher temperature CO phase are no longer energetically favored. The disappearance of the JT distortion yield an accommodation of the $MnO_6$ octahedra in a less distorted structure, compatible with the sudden decrease of the cell volume. Subsequently, charge mobility is restored, which favors the growth of frozen FM clusters which evolve into a long range FM state within the CO background. Thus the structural change precedes the FM one. On the other hand, when the sample temperature is fixed and the field is varied (Fig. 5), the magnetic field promotes the simultaneous occurrence of ferromagnetism and structural contraction, a clear indication of the interplay between the elastic and magnetic properties, similar to that observed in a metallic antiferromagnet, CoMnSi [23]. For temperatures above the pronounced contraction of the sample the magnetic and structural transitions are reversible, and when the field is removed the system returns to the non-FM and highly distorted state. At temperatures below the structural distortion the process is irreversible, and the sample remains in the FM less distorted state after the magnetic field is removed, a characteristic feature of phase separated manganites with magnetic and structural states separated by energy barriers [3,21].

**4 - Conclusions**

To summarize, our results demonstrate that magneto-structural coupling plays a major role in the balance between competing phases in phase separated manganites. Chemical pressure through doping and the application of an external magnetic field have a similar effect, enhancing the FM phase fraction of the compound. Sr doping in the $La_{0.225}Pr_{0.4}(Ca_{1-x}Sr_x)_{0.375}MnO_3$ system increases the cell volume, and at the same time reduces the distortion of the Mn-O-Mn bond angle, which favors the FM double exchange interaction. In one of the low doped compounds ($x$ = 0.05), where the low temperature phase separation is most evident, a contraction of sample's dimension is observed on cooling, at temperatures just above the sudden enhancement of the FM phase fraction. On the other hand, magnetic and structural transitions occurs simultaneously when magnetic field is applied at a fixed temperature. These facts are a clear signal that the FM phase only develops within a low distorted structure. If one forces the magnetic moments to align with an external magnetic field, this forces the structure to adopt a less distorted state. These features occur in a small region of Sr doping: for higher Sr doped samples ($x \geq 0.15$), the larger Mn-O-Mn bond angle and Mn-O distances give rise to a less distorted structure and consequently to a homogeneous FM state. All these facts suggest that microscopic structural inhomogeneities are linked to the formation of the phase separated state in manganites: the emergence of FM regions within the CO phase is driven by local structural changes that favors the double exchange mechanism. Our findings confirm that a sizable ferromagnetic fraction develops even without a magnetic field at temperatures just below a rapid contraction of the structure. When the FM phase is forced by the application of a high magnetic field, the metamagnetic transition occurs simultaneously with the structural changes, a net proof of the correlation between structure and magnetism in phase separated manganites.

## Acknowledgments

This research was supported by the bilateral cooperation FAPERJ-CONICET, the Brazilian agencies CNPq and CAPES (Science without borders program), and by the Brazilian Synchrotron Light Laboratory (LNLS). We acknowledge the useful help of Fabiano Yokaichiya and Helio Salim Amorim with the analysis of the x-ray data. LFC acknowledges the UK EPSRC.

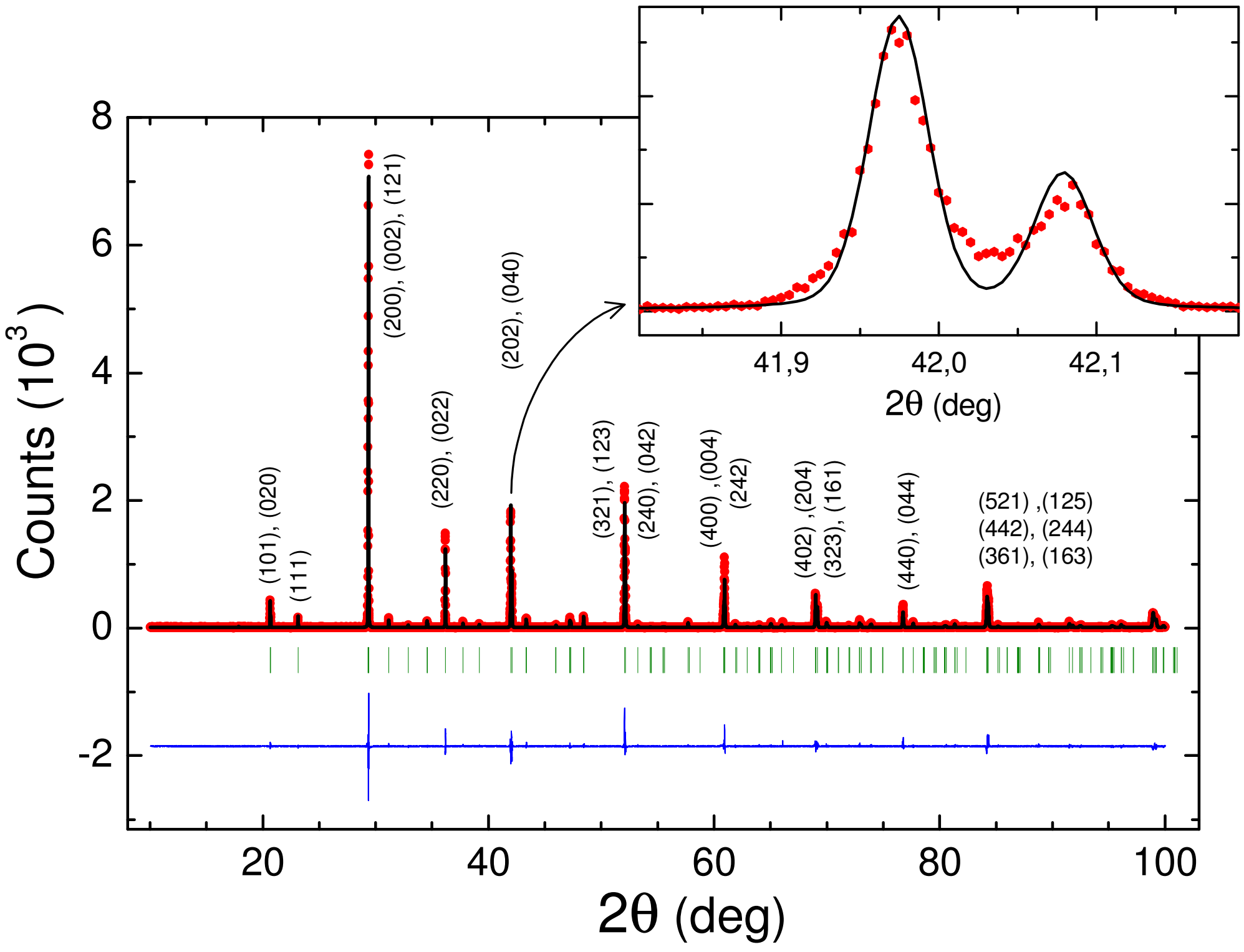

Fig. 1 - Room temperature X-ray diffraction of $La_{0.225}Pr_{0.4}(Ca_{1-x}Sr_x)_{0.375}MnO_3$ , with x=0.05. The symbols are the measured data, the line is the calculated profile, and the difference is shown below. Small vertical bars indicate the expected Bragg peak positions. Miller indices for the main diffraction peaks are given.

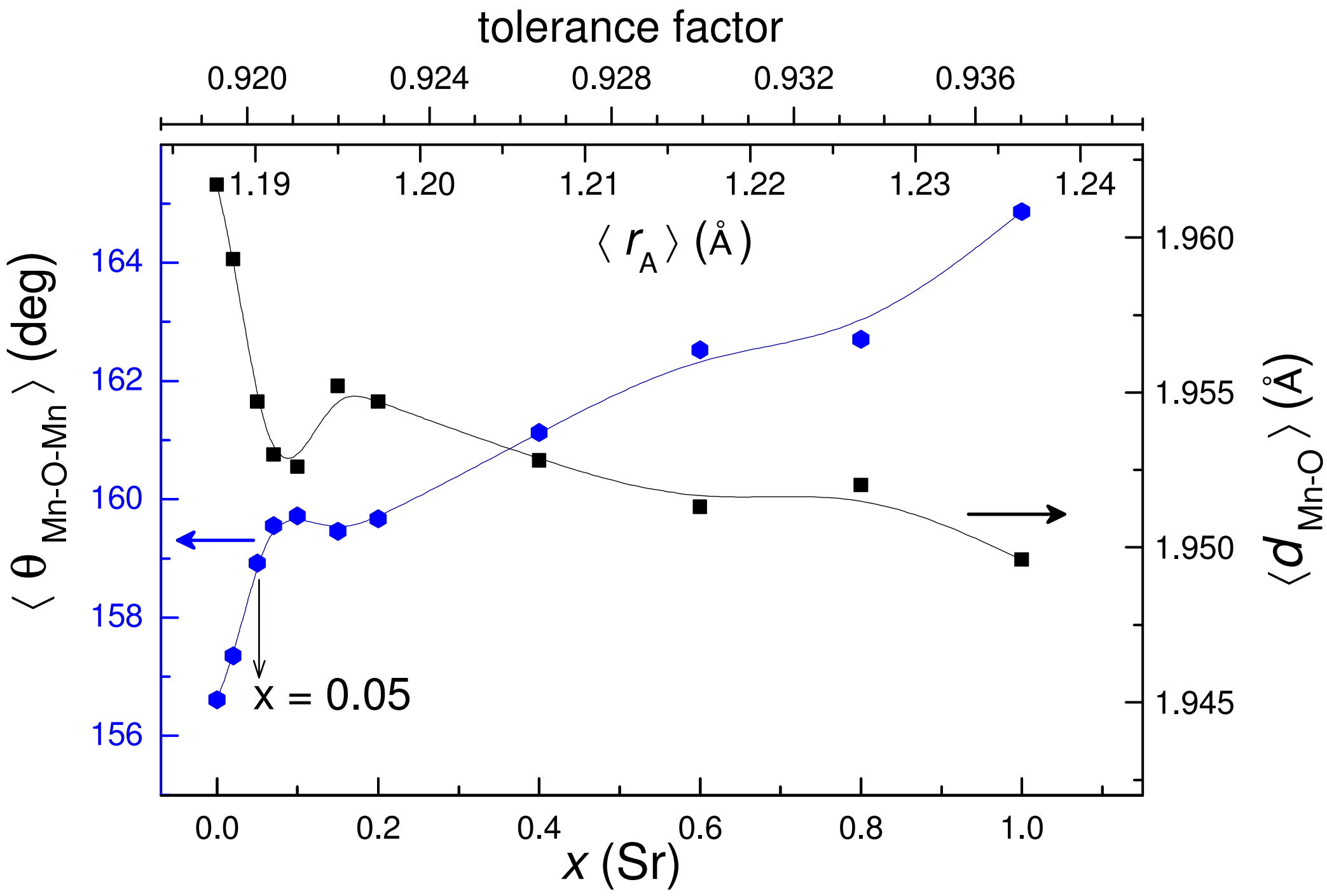

Fig. 2 - Structural parameters of $La_{0.225}Pr_{0.4}(Ca_{1-x}Sr_x)_{0.375}MnO_3$ samples: the Mn-O-Mn bond angle and the Mn-O distance, obtained from Rietveld analysis of the x-ray diffraction measurements as a function of the Sr doping, *x*. Also shown is the correspondence between *x*, the average ionic radius of the A site, $\langle r_A \rangle$, and the tolerance factor.

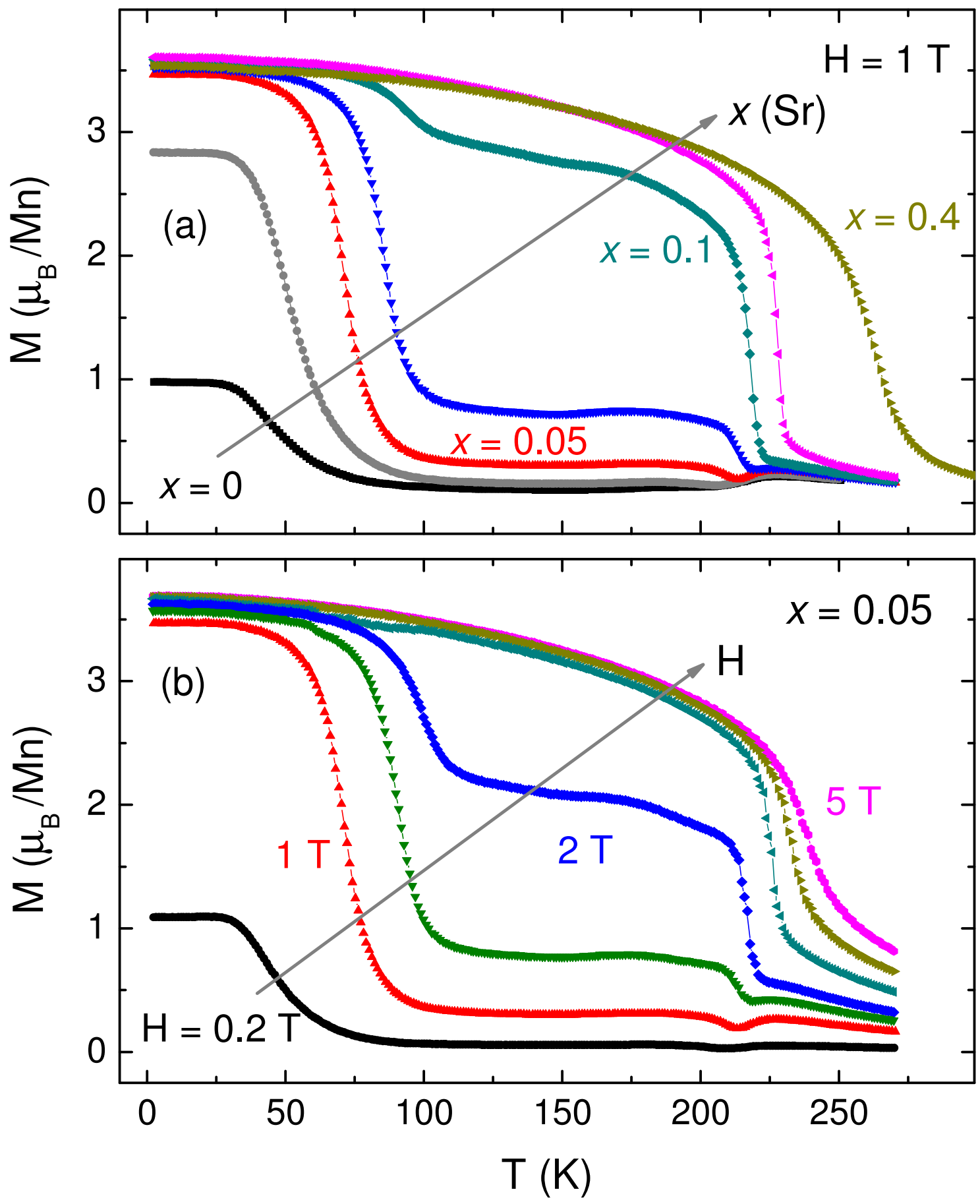

Fig.3 - Field cooled magnetization as a function of temperature for (a) samples of $La_{0.225}Pr_{0.4}(Ca_{1-x}Sr_{x})_{0.375}MnO_3$, with $x$ = 0, 0.02, 0.05, 0.07, 0,10, 0.15, and 0.4, measured with H = 1 T; and (b) for the $x$ = 0.05 sample, measures with H = 0.2 T, 1 T, 1.5 T, 2 T, 3 T, 4 T, and 5 T.

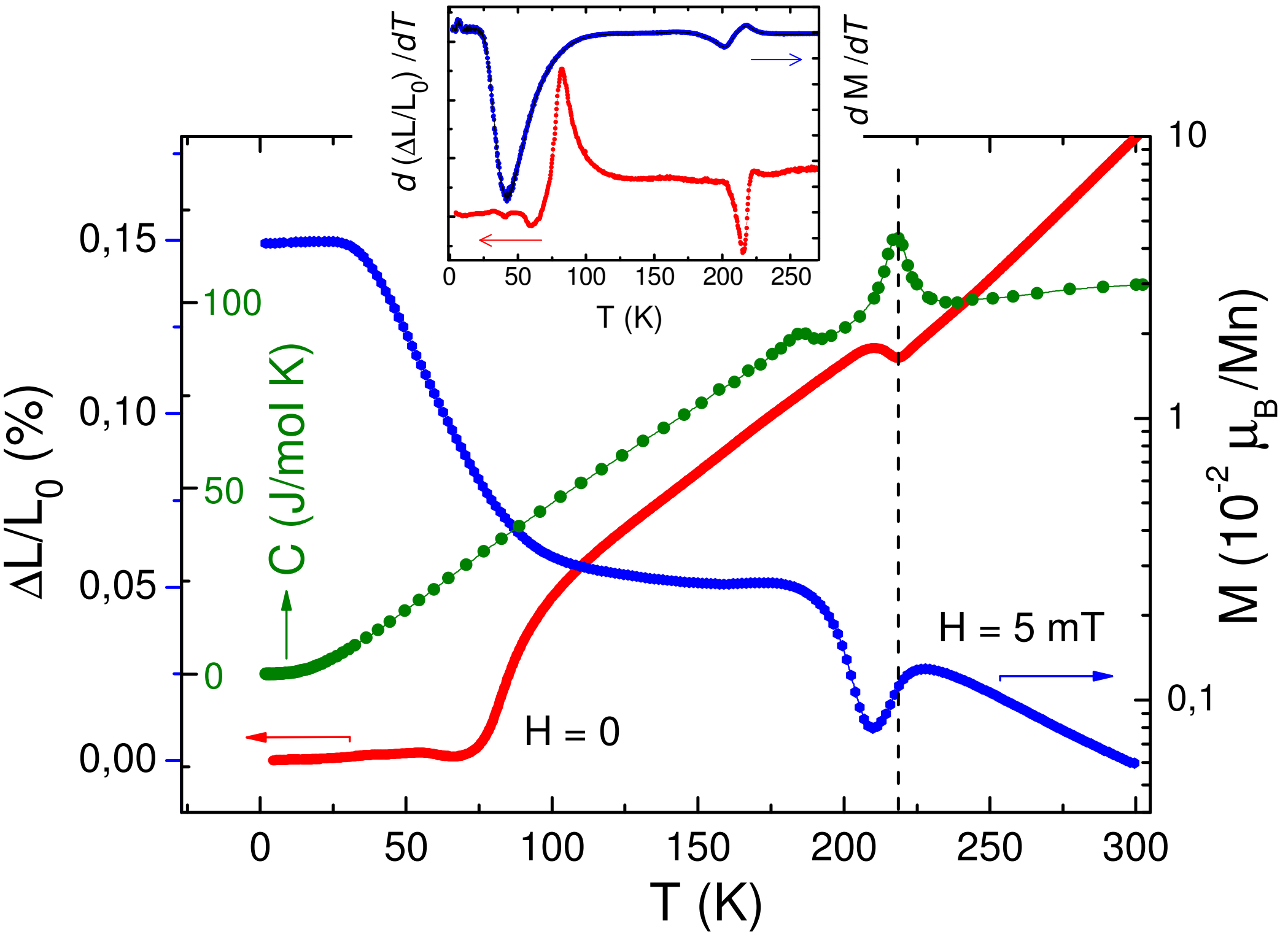

Fig. 4 - Temperature dependence of the relative thermal expansion with H = 0 (red), field cooled logarithmic plot of the magnetization with H = 0.005 T (blue), and specific heat with H = 0 (green), measured on the *x* = 0.05 sample, $La_{0.225}Pr_{0.4}(Ca_{0.95}Sr_{0.05})_{0.375}MnO_3$. The dotted line, at T = 219 K, indicates the CO transition. The inset shows the temperature derivatives of the magnetization and the thermal expansion; the latter is known as the thermal expansion coefficient.

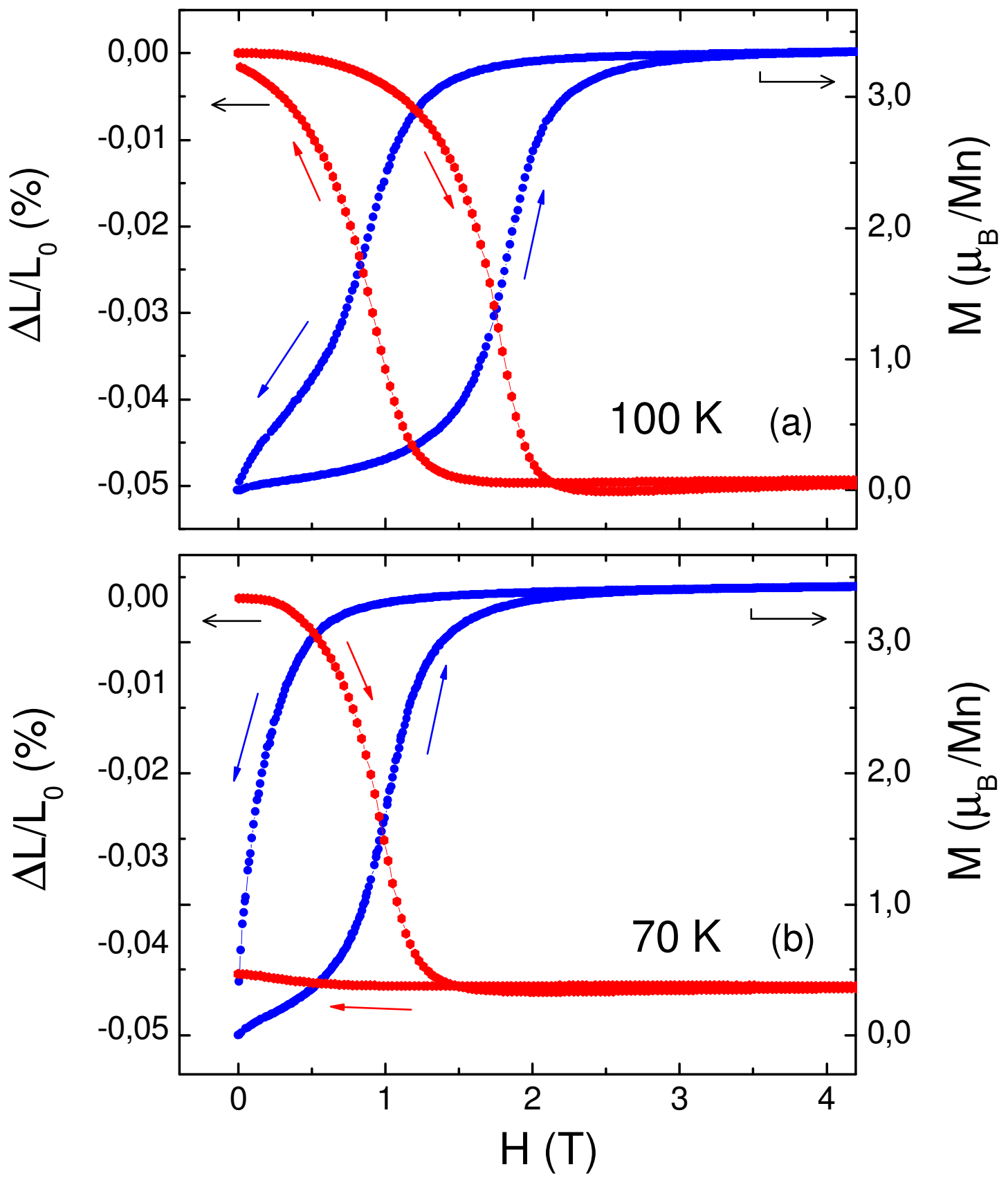

Fig 5 - Magnetostriction and magnetization of La0$_{.225}$$Pr_{0.4}(Ca_{0.95}Sr_{0.05})_{0.375}MnO_3$ , measured at as a function of applied field at 70 and 100 K.